\documentclass[jcp,twocolumn]{revtex4}
\usepackage{amsmath}
\usepackage{color}

\begin{document}
\title{Mode-coupling theory for the steady-state dynamics of active Brownian particles}

\author{Grzegorz Szamel}
\affiliation{Department of Chemistry, Colorado State University, 
Fort Collins, Colorado 80523, USA}

\date{\today}

\begin{abstract}
We present a theory for the steady-state dynamics of a two-dimensional
system of spherically symmetric active Brownian particles. 
The derivation of the theory consists of two steps. First, we integrate out 
the self-propulsions and obtain a many-particle evolution equation
for the probability distribution of the particles' positions. Second, 
we use projection operator technique and a mode-coupling-like 
factorization approximation
to derive an equation of motion for the density correlation function.
The nonequilibrium character of the active system manifests itself through 
the presence of a steady-state correlation function that quantifies spatial 
correlations of microscopic steady-state currents of the particles. 
This function determines the dependence of the short-time dynamics on the
activity. It also enters into the expression for the memory matrix and thus
influences the long-time glassy dynamics. 
\end{abstract}

\maketitle

\section{Introduction}\label{sec:intro}

Recently, there has been a lot of interest in the structure and dynamics of strongly 
interacting active matter systems 
\cite{Ramaswamyrev,Catesrev,Marchettirev,Bechingerrev}. 
This is motivated by a combination of experimental and simulational studies that 
uncovered fascinating phenomena with no analogs in equilibrium thermal (passive) 
systems. Very recent examples include transition from turbulent to coherent flows in
confined three-dimensional active fluids \cite{Dogic2017}, 
spontaneously flowing two-dimensional crystals \cite{Dauchot2018},  
and structure and dynamics of active systems on curved geometries
\cite{Marchetti2017,Janssen2017}.
The motivation comes also from the fact that activity can significantly and
sometimes surprisingly modify phenomena occurring in thermal systems. For
example, early works \cite{Henkes2011,BerthierKurchan,Ni2013,Berthier2014} 
showed that introducing 
active forces changes the glassy behavior of an equilibrium fluid instead of wiping out 
the glassy phase completely, which could have been expected on the basis of a perhaps 
naive analogy with what happens when a shear flow is imposed on an equilibrium colloidal
glass. The resulting non-equilibrium glassy behavior and the eventual glass transition
exhibit features observed in equilibrium supercooled liquids \cite{Berthier2011}, 
including slowing down, transient localization of particles' positions, non-exponential 
relaxations and dynamic heterogeneity. However, also present are unique non-equilibrium
features such as non-trivial equal-time velocity correlations \cite{Szamel2015} and
a variety of different effective temperatures \cite{Levis2015,Preisler2016}. 
It is this confluence of well-known but still not fully understood 
\cite{Berthier2011} glassy behavior and strongly non-equilibrium characteristics
of active matter systems that makes non-equilibrium glassy behavior so fascinating.

Most of the experimental and many of the simulated active glassy systems are quite
complex and are characterized by many independent parameters, which makes systematic 
studies quite involved. For this reason, in our initial investigations of 
non-equilibrium glassy behavior we focused on a a very simple active system, 
the so-called active Ornstein-Uhlenbeck particles (AOUPs) model \cite{Fodor2016} 
introduced independently in Refs. \cite{Szamel2014} and \cite{Maggi2015}. 
In this model, the dynamics is overdamped, there is no thermal noise, and the
particles move under the combined influence of the inter-particle forces and the
self-propulsion. The self-propulsion  
evolves according to the Ornstein-Uhlenbeck process, independently of the
configuration of the particles. For a given interaction potential, the AOUP system 
is characterized by three parameters, the number density $\rho$, the single-particle
effective temperature $T_\text{eff}$ characterizing the driving energy, and the 
persistence time of the self-propulsion $\tau_p$. In the limit of the vanishing 
persistence time, $\tau_p\to 0$, the AOUP system becomes equivalent to a
thermal (passive) system at the temperature equal to the single-particle 
effective temperature. Thus, for the AOUP system the departure from equilibrium 
is characterized by a single parameter, the persistence time of the self-propulsion.

To analyze the dynamics of dense systems of AOUPs we used a combination of 
simulations and analytical theory. In collaboration with E. Flenner and L. Berthier,
we uncovered the significance of the equal-time, steady-state correlation function
of particles' velocities \cite{Szamel2015}. 
This function exhibits strong wavevector dependence
for long persistence times and becomes trivial, \textit{i.e.} wavevector
independent and related to the temperature, in the 
equilibrium limit $\tau_p\to 0$. Furthermore, we showed that increasing departure 
from equilibrium can result in both faster dynamics and fluidization of 
a glassy system \cite{Berthier2017}, which was also observed by other 
workers \cite{Ni2013,Berthier2014,Mandal2014}, and, unexpectedly, 
slower dynamics and glassification of 
a fluid system \cite{Flenner2016,Berthier2017}. We found that this highly non-trivial 
behavior of the effective glass transition line in the $T_\text{eff}-\rho$ plane 
correlates with the dependence of the steady-state structure on the
departure from equilibrium \cite{Berthier2017}.

On the theory side, we derived an approximate theory for the steady-state dynamics
of AOUPs \cite{Szamel2016}. The basic assumption of this theory was the absence of
the steady state currents, or more precisely, that 
the velocities of individual particles vanish after averaging over the 
self-propulsions. The derivation was done in two steps. First, we approximately
integrated out the self-propulsions and obtained an effective equation of motion for 
the many-particle distribution of particles' positions. This equation featured 
a time-dependent diffusivity matrix. The time-dependence was crucial for retaining 
proper (ballistic) short-time dynamics of the AOUPs. 
Next, using the effective many-particle equation of motion we derived an approximate
equation of motion for the intermediate scattering function. In this last step
we used a factorization approximation analogous to that employed in the 
mode-coupling theory of glassy dynamics and the glass transition \cite{Goetzebook}.
The resulting self-consistent equation of motion for the intermediate scattering 
function resembled an equation of motion for an under-damped colloidal
system (without hydrodynamic interactions). This could have been expected since 
the AOUP dynamics is ballistic at short times and diffusive at long times, as in an 
under-damped colloidal system. However, we found that the equation of motion for the 
intermediate scattering function depends, in a highly non-trivial way, on the activity
of the system. This dependence manifests itself through
the correlation function of particles' velocities, which entered into an
analogue of the frequency matrix term (and thus determined the short-time dynamics)
and into the vertices of the approximate expression for the memory function.

In the present contribution we extend the derivation presented in Ref. \cite{Szamel2016}
to the most often studied active system, the so-called active Brownian
particles (ABPs) model \cite{tenHagen,FilyMarchetti}. 
In this model, the dynamics is overdamped but there
is also thermal noise. Thus, the particles move under the combined influence of 
the inter-particle forces, thermal noise originating from fluctuations of
the solvent, and the self-propulsion. The magnitude of the self-propulsion is
fixed and its direction changes via rotational diffusion. For a given interaction
potential, the ABP system is characterized by four parameters, the number density 
$\rho$, the translational diffusion coefficient $D_t$, which depends in
the temperature $T$ characterizing the thermal noise, the magnitude
of the self-propulsion $v_0$ and the rotational diffusion constant specifying 
the evolution of the direction of the self-propulsion $D_r$. For real active 
colloidal particles both $D_t$ and $D_r$ are determined by the hydrodynamics
of the solvent and are, therefore, related but we will treat them as independent
model parameters, as was done in some simulational \cite{Ni2013} and 
theoretical \cite{Voigtmann2017} studies.  

Compared to the AOUP system, the ABP model introduces two complications. 
First, we need to include thermal noise (translational diffusion) in addition 
to the self-propulsion. Second, in the ABP model the relation between 
the variable describing the state of the self-propulsion and the particle
motion is non-linear and thus even the solution of the single-particle motion
is highly non-trivial \cite{Franosch2016}. 

To put our work in the context of earlier microscopic theoretical investigations we
would like to mention two other theories for the dynamics of the ABP system 
\cite{Voigtmann2017,Farage} and a theory for a generalized AOUP model that
includes thermal noise \cite{Feng2017}. All these theoretical approaches share
one important feature with our theory for the AOUP system \cite{Szamel2016} 
and with the theory presented here: all these theories rely upon a factorization 
approximation and thus are of mode-coupling flavor. However, they differ in 
their focus and in the way they include the self-propulsion. In the next 
two paragraphs we will delineate these differences. We shall also mention 
a theory of Nandi and Gov \cite{NandiGov2017}, which falls in between 
microscopic theories of Refs. \cite{Voigtmann2017,Farage,Feng2017} and 
the theoretical analysis of an effective $p$-spin-like active model of 
Ref. \cite{BerthierKurchan}.

The difference between the theories of Farage and Brader \cite{Farage} and 
of Liluashvili \textit{et al.} \cite{Voigtmann2017} and both our theories and
the approach of Feng and Hou \cite{Feng2017} is that the former theories 
follow the philosophy of the ``integration-through-transients'' approach
to the dynamics of colloidal systems under shear developed by Fuchs and Cates 
\cite{Fuchsrev}. In this approach one assumes that the system was in 
an equilibrium state in the infinitely distant past and then the drive, in this
case the activity, was turned on. The advantage of this approach is that, 
in principle, it allows one to calculate both equal time and dynamic properties 
of an active system within a single theory. Within this
approach one easily derives approximate equations of motion for transient 
correlation functions. However, it is much more difficult to obtain equations of motion 
for time-dependent correlation functions in a steady state \cite{Krueger2009}.
In addition, it is not clear how one could use this approach to describe a 
fully athermal system that does not have any dynamics without the activity. 
In contrast, our theories and the theory of Feng and Hou focus on the steady-state 
dynamics of the active system. Their disadvantage is that a separate investigation,
either analytical or simulational, is needed to supply these theories with
necessary equal-time steady-state correlation functions. 

On the other hand, the difference between theories of Farage and Brader \cite{Farage}
and Feng and Hou \cite{Feng2017} and both our theories and the theory of 
Liluashvili \textit{et al.} \cite{Voigtmann2017} is in the way the self-propulsion
is treated. In the former theories the active system is approximately replaced by 
a thermal (passive) system using either a very simple time coarse-graining
procedure \cite{Farage} or a procedure introduced by Fox \cite{Fox} in
the context of approximating a system with a colored noise by an equivalent
system with white noise. In contrast, both our theories and the theory 
of Liluashvili \textit{et al.} include, in an approximate way, the time
evolution of the self-propulsion and the resulting time-delayed response. 
In the theory of Ref. \cite{Szamel2016} and in the theory presented here 
this is achieved by retaining the time dependence in the 
many-particle diffusivity matrix. The theory of Liluashvili \textit{et al.}
treats the self-propulsion explicitly and thus constitutes  the most accurate 
description of the self-propulsion.

The paper is organized as follows. In the next section, Sec. \ref{sec:ABPmodel},
we introduce and briefly discuss the ABP system. 
In Sec. \ref{sec:assumpt} we discuss the basic assumption of our theory and 
its physical meaning. In Sec. \ref{sec:mct} we present the derivation of the 
approximate theory for the long-time dynamics of the intermediate scattering function
of the ABP model. We end with discussion in Sec. \ref{sec:disc}. 

\section{Active Brownian particles}\label{sec:ABPmodel}

We consider a two-dimensional \cite{comment2d} system of $N$ interacting, 
self-propelled particles in a volume (area) $A$. 
The average density $\rho=N/A$. The particles interact via a spherically
symmetric potential $V(r)$. They move in a viscous medium that is characterized
by the friction coefficient of a single particle, which we denote by $\xi_0$. 
We assume that the friction felt by a particle
is independent of the particle density and configuration, and thus we neglect 
hydrodynamic interactions \cite{Dhont}. 
Each particle moves under the combined influence of interparticle force
derived from the potential $V(r)$, random, thermal
forces due to the fluctuations of the solvent and self-propulsion \cite{commentSP}.
We assume that the velocity due to the self-propulsion has a constant magnitude 
$v_0$ and that it evolves in time via rotational diffusion. 
The corresponding equations of motion read,
\begin{eqnarray}
\label{emotion1}
\dot{\mathbf{r}}_i & = & \xi_0^{-1}\left[ \mathbf{F}_i + \boldsymbol{\eta}_i \right]
+ v_0 \mathbf{e}(\theta_i), \\
\label{emotion2}
\dot{\theta}_i & =  &  \eta_{\theta i}.
\end{eqnarray}
In Eq.~\eqref{emotion1}, $\mathbf{r}_i$ is the position of particle $i$ 
and $\mathbf{F}_i$ is the force acting on particle $i$ originating from the interactions,
\begin{equation}
\mathbf{F}_i = -\sum_{j\neq i} \boldsymbol{\nabla}_i V(r_{ij}),
\end{equation}   
where $\boldsymbol{\nabla}_i$ denotes a partial derivative 
with respect to $\mathbf{r}_i$, 
$\boldsymbol{\nabla}_i = \frac{\partial}{\partial \mathbf{r}_i}$.
Furthermore, in Eq.~\eqref{emotion1},
$\boldsymbol{\eta}_i$ is the Gaussian thermal noise with zero mean and variance
$\left<\boldsymbol{\eta}_i(t)\boldsymbol{\eta}_j(t^\prime)\right> =
2 \xi_0 T \mathbf{I} \delta_{ij}\delta(t-t^\prime)$ with $T$ being the temperature
(we use units such that the Boltzmann constant $k_B=1$) 
and $\mathbf{I}$ being the unit tensor. Finally, in Eq.~\eqref{emotion1}, 
$\mathbf{e}(\theta_i)$ is the unit vector specifying the direction of the 
self-propulsion, $\mathbf{e}(\theta_i) 
= (\cos\theta_i,\sin\theta_i)$. The single particle translational diffusion coefficient
is proportional to the temperature, $D_t=T/\xi_0$.  
In Eq.~\eqref{emotion2}, $\eta_{\theta i}$
is an internal Gaussian noise with zero mean and variance
$\left<\eta_{\theta i}(t)\eta_{\theta j}(t^\prime)\right> =
2 D_r \delta_{ij}\delta(t-t^\prime)$, with $D_r$ being the rotational
diffusion coefficient. As mentioned earlier, we treat $D_t$ and $D_r$ as 
independent parameters. 

We emphasize that the direction of each self-propulsion evolves 
independently of positions of the particles and of (direct) forces acting on the 
particles. 

Without interactions, particles evolving according to
Eqs. (\ref{emotion1}-\ref{emotion2}) perform a somewhat complicated random 
walk, with non-trivial higher-order cumulants, but with a relatively simple 
mean-square displacement \cite{tenHagen}
\begin{equation}\label{msdfree}
\left<\left(\mathbf{r}_i(t)-\mathbf{r}_i(0)\right)^2\right> = 
4 D_t t + 2 \frac{v_0^2}{D_r^2}\left(D_r t -1 + e^{-D_r t}\right).
\end{equation}
One should note that the expression for the mean-square displacement is 
the same as for a simpler system of AOUPs with additional thermal noise. 
According to Eq. (\ref{msdfree}), the self-propulsion contributes to the 
mean-square displacement at the level of $t^2$,
\begin{equation}\label{msdfreess}
\left<\left(\mathbf{r}_i(t)-\mathbf{r}_i(0)\right)^2\right> \approx
4 D_t t + \frac{v_0^2}{D_r^2} t^2 \;\;\;\;\; t\ll D_r^{-1}
\end{equation}
and the long-time motion, $t\gg D_r^{-1}$, is diffusive with diffusion coefficient 
$D_\text{eff}$,
\begin{equation}\label{difffree}
D_\text{eff} = D_t + \frac{v_0^2}{2 D_r}
\end{equation}
Comparing expression (\ref{difffree}) with the well-known formula for the 
diffusion coefficient of a Brownian particle moving in a viscous medium with 
friction constant $\xi_0$, $D_\text{Brownian} = T/\xi_0$, we can 
define the \textit{single-particle} effective temperature,
\begin{equation}\label{Teff}
T_{\text{eff}}= D_\text{eff} \xi_0  = T + \frac{v_0^2}{2 D_r\xi_0}.
\end{equation}

We note that in the limit of fast evolution of the self-propulsions,
$D_r\to\infty$, $v_0\to\infty$, $v_0^2/D_r=\text{ const.}$, the mean-square
displacement as given by Eq. (\ref{msdfree}) becomes linear in time. In fact,
in this limit the system becomes equivalent to a thermal system with the 
temperature given by Eq. (\ref{Teff}).

For further theoretical analysis it is convenient to replace the description
of the system's dynamics in terms of the equations of motion 
(\ref{emotion1}-\ref{emotion2}) by the equivalent description
in terms of an $N$-particle joint probability distribution
of positions and self-propulsions  
$P_N(\mathbf{r}_1,\mathbf{e}_1,...,\mathbf{r}_N,\mathbf{e}_N;t)
\equiv P_N(\mathbf{r}_1,\theta_1,...,\mathbf{r}_N,\theta_N;t)$. This distribution
evolves in time with evolution operator $\Omega$,
\begin{equation}\label{eom}
\partial_t P_N(\mathbf{r}_1,\theta_1,...,\mathbf{r}_N,\theta_N;t) 
= \Omega P_N(\mathbf{r}_1,\theta_1,...,\mathbf{r}_N,\theta_N;t),
\end{equation}
which can be derived from equations of motion (\ref{emotion1}-\ref{emotion2}),
\begin{equation}\label{Omega1}
\Omega = D_t \sum_{i=1}^{N} 
\boldsymbol{\nabla}_i\cdot\left[\boldsymbol{\nabla}_i - \beta\mathbf{F}_i\right] 
- \sum_{i=1}^{N} v_0 \boldsymbol{\nabla}_i\cdot\mathbf{e}(\theta_i)
+D_r \sum_i \partial^2_{\theta_i}.
\end{equation} 

We assume that there exists a steady state. In other words, we assume there
exists a probability distribution 
$P_N^{ss}(\mathbf{r}_1,\theta_1,...,\mathbf{r}_N,\theta_N)$ such
that
\begin{equation}\label{ss}
\Omega P_N^{ss}(\mathbf{r}_1,\theta_1,...,\mathbf{r}_N,\theta_N) = 0.
\end{equation}
We expect that there are correlations between positions and self-propulsion 
velocities \cite{BialkeEPL} and thus 
\begin{equation}
P_N^{\text{ss}}(\mathbf{r}_1,\theta_1,...,\mathbf{r}_N,\theta_N)
\neq
P_N^{\text{ss}}(\mathbf{r}_1,...,\mathbf{r}_N)
P_N^{\text{ss}}(\theta_1,...,\theta_N),
\end{equation}
where $P_N^{\text{ss}}(\mathbf{r}_1,...,\mathbf{r}_N)$
and $P_N^{\text{ss}}(\theta_1,...,\theta_N)$ are the steady-state 
distributions of positions and self-propulsions,
\begin{eqnarray}
P_N^{\text{ss}}(\mathbf{r}_1,...,\mathbf{r}_N) \!\!\! &=& \!\!\! 
\int  d\theta_1 ... d\theta_N 
P_N^{\text{ss}}(\mathbf{r}_1,\theta_1,...,\mathbf{r}_N,\theta_N),
\\
P_N^{\text{ss}}(\theta_1,...,\theta_N) \!\!\! &=& \!\!\! 
\int  d\mathbf{r}_1 ... d\mathbf{r}_N 
P_N^{\text{ss}}(\mathbf{r}_1,\theta_1,...,\mathbf{r}_N,\theta_N).
\end{eqnarray}
In general, neither the joint steady-state distribution 
$P_N^{\text{ss}}(\mathbf{r}_1,\theta_1,...,\mathbf{r}_N,\theta_N)$
nor the steady-state distributions of positions
$P_N^{\text{ss}}(\mathbf{r}_1,...,\mathbf{r}_N)$ are known exactly (for approximate
theories for the latter distribution 
see Refs.~\cite{Maggi2015,FarageKB}). However, the steady-state 
distribution of self-propulsions has a trivial form,
\begin{equation}
P_N^{\text{ss}}(\theta_1,...,\theta_N) = \left(2\pi\right)^{-N}.
\end{equation}

The main object of our theory is the intermediate scattering function,
$F(q;t)$, which describes the time dependence of the collective density fluctuations,
\begin{equation}\label{Fqt0}
F(q;t) = \frac{1}{N}
\left<\sum_i e^{-i\mathbf{q}\cdot\mathbf{r}_i(t)}
\sum_j e^{i\mathbf{q}\cdot\mathbf{r}_j(0)} \right>.
\end{equation}
Here and in the following the brackets $\left< \dots \right>$ denote averaging
over a steady-state distribution of positions and self-propulsions.

Evolution operator (\ref{Omega1}) allows us to rewrite the definition of
the intermediate scattering function (\ref{Fqt0}),
\begin{equation}\label{Fqt}
F(q;t) = \frac{1}{N}
\left<n(\mathbf{q}) \exp\left(\Omega t\right) n(-\mathbf{q})\right>.
\end{equation}
In Eq. (\ref{Fqt}) $n(\mathbf{q})$  
is the Fourier transform of the microscopic density,
\begin{equation}\label{n1def}
n(\mathbf{q}) = \sum_l e^{-i\mathbf{q}\cdot\mathbf{r}_l},
\end{equation}
We emphasize that in 
Eq. (\ref{Fqt}) and in all similar formulas the steady-state distribution 
stands to the right of the quantity being averaged, and all operators act on it too.

\section{The main assumption: absence of average currents}\label{sec:assumpt}

We follow our earlier theory for AOUPs \cite{Szamel2016} 
and assume that in the steady state, the currents vanish after integrating out
the self-propulsions. This assumption will allow us to approximate our system by a 
passive system with a time-dependent diffusivity matrix. 

To make our assumption explicit we first rewrite the equation of motion 
for the joint probability distribution
of positions and self-propulsions, Eq. (\ref{eom}), in the form of a continuity
equation,
\begin{eqnarray}\label{cont}
\lefteqn{ 
\partial_t P_N(\mathbf{r}_1,\theta_1,...,\mathbf{r}_N,\theta_N;t) = }
\\ \nonumber && 
-\sum_{i} \boldsymbol{\nabla}_i\cdot 
\mathbf{j}_i(\mathbf{r}_1,\theta_1,...,\mathbf{r}_N,\theta_N;t)
\\ \nonumber && 
-\sum_{i} \partial_{\theta_i} 
j_i^{\theta}(\mathbf{r}_1,\theta_1,...,\mathbf{r}_N,\theta_N;t),
\end{eqnarray}
where current densities are defined as
\begin{eqnarray}\label{currentpos0}
\lefteqn{
\mathbf{j}_i(\mathbf{r}_1,\theta_1,...,\mathbf{r}_N,\theta_N;t) = }
\\ \nonumber && 
\left(-D_t\boldsymbol{\nabla}_i + \xi_0^{-1}\mathbf{F}_i + v_0 \mathbf{e}_i \right)
P_N(\mathbf{r}_1,\theta_1,...,\mathbf{r}_N,\theta_N;t),
\end{eqnarray}
\begin{eqnarray}\label{currentsp0}
\lefteqn{
j_i^{\theta}(\mathbf{r}_1,\theta_1,...,\mathbf{r}_N,\theta_N;t) = }
\\ \nonumber && 
-D_r\partial_{\theta_i}
P_N(\mathbf{r}_1,\theta_1,...,\mathbf{r}_N,\theta_N;t).
\end{eqnarray}

Current densities (\ref{currentpos0}-\ref{currentsp0}) are microscopic
quantities (\textit{i.e} in principle they depend on positions and self-propulsions
of all the particles), which may be non-zero in a system without detailed balance. 
We assume that in the steady state, 
the current density in the position space integrated over self-propulsions  
vanishes,
\begin{eqnarray}\label{curvan}
&& \mathbf{j}_i^{\text{ss}}(\mathbf{r}_1,...,\mathbf{r}_N;t) = 
\int  d\theta_1 ... d\theta_N 
\mathbf{j}_i^{\text{ss}}(\mathbf{r}_1,\theta_1,...,\mathbf{r}_N,\theta_N;t)
\nonumber \\ && \equiv 
\int  d\theta_1 ... d\theta_N
\left(-D_t\boldsymbol{\nabla}_i + \xi_0^{-1}\mathbf{F}_i + v_0 \mathbf{e}_i \right)
\nonumber \\ && \times
P_N^{\text{ss}}(\mathbf{r}_1,\theta_1,...,\mathbf{r}_N,\theta_N)
= 0.
\end{eqnarray}

Assumption (\ref{curvan}) implies the following expression for the local 
steady-state average of the self-propulsion
\begin{equation}\label{avsp}
v_0\left<\mathbf{e}_i\right>_{\text{lss}} = - \xi_0^{-1} \mathbf{F}_i 
+ D_t \boldsymbol{\nabla}_i \ln P_N^{\text{ss}}(\mathbf{r}_1,...,\mathbf{r}_N)
\end{equation}
where
the local steady-state average is defined as
\begin{eqnarray}
\lefteqn{ \left< \dots \right>_{\text{lss}} = }
\\ \nonumber && 
\frac{1}{P_N^{\text{ss}}(\mathbf{r}_1,...,\mathbf{r}_N)}
\int  d\theta_1 ... d\theta_N
\dots
P_N^{\text{ss}}(\mathbf{r}_1,\theta_1,...,\mathbf{r}_N,\theta_N).
\end{eqnarray}
Eq. (\ref{avsp}) expresses a balance of the self-propulsion
acting on particle $i$ and the sum of the total potential force acting on this particle
and the averaged force due to the solvent fluctuations, for a 
given configuration, \textit{i.e.} for a given set of the positions of the 
particles.  

We are  not aware of any study that specifically focused on the existence
of non-trivial steady-state currents in high density active systems without aligning 
interactions \cite{commentcurrents}. 
We note that the assumption (\ref{curvan}) is made at the level of $N$-particle 
quantities. Thus, its direct simulational verification seems rather difficult.
However, it might be possible to define and measure reduced (few-particle) current 
densities. Work in this direction is planned and the results will be reported
in the future. 

\section{Mode-coupling theory}\label{sec:mct}

\subsection{Effective evolution operator for particles' positions}

We again follow our earlier theory \cite{Szamel2016} and start the analysis of the 
ABP system dynamics by deriving 
an approximate equation of motion for the $N$-particle distribution of particles' 
positions. We note that since we retain the time-dependence that originates from the 
evolution of particles' self-propulsions, this step is exact.
The reason for introducing approximations is to make the formal expressions explicit.
Thus, in principle, we are not restricted to fast evolution of self-propulsions. 
Only if we neglect the time delay caused by evolution of self-propulsions on finite 
time scale and introduce an effective passive system with Markovian dynamics we are 
assuming that the evolution of self-propulsions is fast compared to the 
evolution of particles' positions. 

We start by introducing a projection operator that acts on
an $N$-particle probability distribution of self-propulsions and positions and
projects it on a local steady-state distribution, \textit{i.e.} on a distribution 
in which self-propulsions have a steady-state distribution for a given sample 
of positions,
\begin{widetext}
\begin{eqnarray}
\mathcal{P}_{\text{lss}} 
P_N(\mathbf{r}_1,\theta_1,...,\mathbf{r}_N,\theta_N;t)
&=& 
\frac{P_N^{\text{ss}}(\mathbf{r}_1,\theta_1,...,\mathbf{r}_N,\theta_N)}
{P_N^{\text{ss}}(\mathbf{r}_1,...,\mathbf{r}_N)}
\int  d\theta_1 ... d\theta_N
P_N(\mathbf{r}_1,\theta_1,...,\mathbf{r}_N,\theta_N;t)
\nonumber \\ 
&=& \frac{P_N^{\text{ss}}(\mathbf{r}_1,\theta_1,...,\mathbf{r}_N,\theta_N)}
{P_N^{\text{ss}}(\mathbf{r}_1,...,\mathbf{r}_N)} 
P_N(\mathbf{r}_1,...,\mathbf{r}_N;t).
\end{eqnarray}
\end{widetext}
We note that by integrating $ \mathcal{P}_{\text{lss}} 
P_N(\mathbf{r}_1,\theta_1,...,\mathbf{r}_N,\theta_N;t)$ over self-propulsions
we get the probability distribution of particles' positions, 
$P_N(\mathbf{r}_1,...,\mathbf{r}_N;t)$.

Next, we define the orthogonal projection,
\begin{eqnarray}
\mathcal{Q}_{\text{lss}} = \mathcal{I} - \mathcal{P}_{\text{lss}},
\end{eqnarray}
and write down equations of motion for 
$\mathcal{P}_{\text{lss}} 
P_N(\mathbf{r}_1,\theta_1,...,\mathbf{r}_N,\theta_N;t)$ and
$\mathcal{Q}_{\text{lss}} 
P_N(\mathbf{r}_1,\theta_1,...,\mathbf{r}_N,\theta_N;t)$,
\begin{eqnarray}\label{PPeom}
\lefteqn{
\partial_t \mathcal{P}_{\text{lss}} 
P_N(\mathbf{r}_1,\theta_1,...,\mathbf{r}_N,\theta_N;t)=}
\nonumber \\ && 
\mathcal{P}_{\text{lss}} \Omega \mathcal{P}_{\text{lss}} 
P_N(\mathbf{r}_1,\theta_1,...,\mathbf{r}_N,\theta_N;t) 
\nonumber \\ && + 
\mathcal{P}_{\text{lss}} \Omega \mathcal{Q}_{\text{lss}} 
P_N(\mathbf{r}_1,\theta_1,...,\mathbf{r}_N,\theta_N;t),
\end{eqnarray}
\begin{eqnarray}\label{QPeom}
\lefteqn{
\partial_t \mathcal{Q}_{\text{lss}} 
P_N(\mathbf{r}_1,\theta_1,...,\mathbf{r}_N,\theta_N;t) = }
\nonumber \\ && 
\mathcal{Q}_{\text{lss}} \Omega \mathcal{P}_{\text{lss}} 
P_N(\mathbf{r}_1,\theta_1,...,\mathbf{r}_N,\theta_N;t) 
\nonumber \\ && 
+ \mathcal{Q}_{\text{lss}} \Omega \mathcal{Q}_{\text{lss}} 
P_N(\mathbf{r}_1,\theta_1,...,\mathbf{r}_N,\theta_N;t).
\end{eqnarray}

Since our goal is to calculate the intermediate scattering function,
Eq. (\ref{Fqt}), which is a function of positions only, we can restrict ourselves
to initial states that satisfy the following condition, 
\begin{equation}
\mathcal{Q}_{\text{lss}}
P_N(\mathbf{r}_1,\theta_1,...,\mathbf{r}_N,\theta_N;t=0)=0.
\end{equation} 
Then we solve Eqs. (\ref{PPeom}-\ref{QPeom}) for 
the Laplace transform, $\mathcal{LT}$, of 
$\partial_t \mathcal{P}_{\text{lss}} 
P_N(\mathbf{r}_1,\theta_1,...,\mathbf{r}_N,\theta_N;t)$ and we obtain
\begin{widetext}
\begin{eqnarray}\label{proj1}
\mathcal{LT}\left[\partial_t \mathcal{P}_{\text{lss}} 
P_N(\mathbf{r}_1,\theta_1,...,\mathbf{r}_N,\theta_N;t)\right](z) = 
\left[  \mathcal{P}_{\text{lss}} \Omega \mathcal{P}_{\text{lss}} 
+ \mathcal{P}_{\text{lss}} \Omega \mathcal{Q}_{\text{lss}}
\frac{1}{z-\mathcal{Q}_{\text{lss}} \Omega \mathcal{Q}_{\text{lss}}} 
\mathcal{Q}_{\text{lss}}\Omega\mathcal{P}_{\text{lss}} \right]
\mathcal{P}_{\text{lss}} P_N(\mathbf{r}_1,\theta_1,...,\mathbf{r}_N,\theta_N;z).
\end{eqnarray}
The first term inside the brackets on right-hand-side of Eq. (\ref{proj1}) reads
\begin{eqnarray}\label{proj1firstterm}
\lefteqn{ \mathcal{P}_{\text{lss}} \Omega \mathcal{P}_{\text{lss}} 
P_N(\mathbf{r}_1,\theta_1,...,\mathbf{r}_N,\theta_N;z) }
\nonumber \\ && 
= - \frac{P_N^{\text{ss}}(\mathbf{r}_1,\theta_1,...,\mathbf{r}_N,\theta_N)}
{P_N^{\text{ss}}(\mathbf{r}_1,...,\mathbf{r}_N)}
\int  d\theta_1 ... d\theta_N \sum_i \left\{ 
\left[-D_t \boldsymbol{\nabla}_i+ \xi_0^{-1}\mathbf{F}_i + v_0 \mathbf{e}_i \right]
P_N^{\text{ss}}(\mathbf{r}_1,\theta_1,...,\mathbf{r}_N,\theta_N)\right\}\cdot
\boldsymbol{\nabla}_i \frac{P_N(\mathbf{r}_1,...,\mathbf{r}_N;z)}
{P_N^{\text{ss}}(\mathbf{r}_1,...,\mathbf{r}_N)}
\nonumber \\ &&
+ \frac{P_N^{\text{ss}}(\mathbf{r}_1,\theta_1,...,\mathbf{r}_N,\theta_N)}
{P_N^{\text{ss}}(\mathbf{r}_1,...,\mathbf{r}_N)}
\int  d\theta_1 ... d\theta_N \sum_i
P_N^{\text{ss}}(\mathbf{r}_1,\theta_1,...,\mathbf{r}_N,\theta_N) 
D_t \boldsymbol{\nabla}_i\cdot
\boldsymbol{\nabla}_i \frac{P_N(\mathbf{r}_1,...,\mathbf{r}_N;z)}
{P_N^{\text{ss}}(\mathbf{r}_1,...,\mathbf{r}_N)}
\end{eqnarray}
We see that if current densities vanish in the steady state, 
Eq. (\ref{curvan}), the first term at the right-hand side of Eq. (\ref{proj1firstterm})
vanishes and the second term can be re-written in terms of the 
effective steady state force,
\begin{eqnarray}\label{proj1firstterm2}
\lefteqn{ \mathcal{P}_{\text{lss}} \Omega \mathcal{P}_{\text{lss}} 
P_N(\mathbf{r}_1,\theta_1,...,\mathbf{r}_N,\theta_N;z) }
\nonumber \\ && 
= \frac{P_N^{\text{ss}}(\mathbf{r}_1,\theta_1,...,\mathbf{r}_N,\theta_N)}
{P_N^{\text{ss}}(\mathbf{r}_1,...,\mathbf{r}_N)}
\sum_i D_t \boldsymbol{\nabla}_i\cdot
\left\{\boldsymbol{\nabla}_i 
- \left[\boldsymbol{\nabla}_i \ln P_N^{\text{ss}}(\mathbf{r}_1,...,\mathbf{r}_N) \right]
\right\}
P_N(\mathbf{r}_1,...,\mathbf{r}_N;z).
\end{eqnarray}
Furthermore, combining the assumption that currents vanish with Eq. (\ref{avsp}) 
one can show that
\begin{eqnarray}\label{right}
\mathcal{Q}_{\text{lss}}\Omega \mathcal{P}_{\text{lss}} P_N(z) &=&
- \sum_i 
\left(- D_t \boldsymbol{\nabla}_i \ln P_N^{\text{ss}}(\mathbf{r}_1,...,\mathbf{r}_N)
+ \xi_0^{-1} \mathbf{F}_i + v_0\mathbf{e}_i \right)
P_N^{\text{ss}}(\mathbf{r}_1,\theta_1,...,\mathbf{r}_N,\theta_N)
\cdot
\left[\boldsymbol{\nabla}_i \frac{P_N(\mathbf{r}_1,...,\mathbf{r}_N;z)}
{P_N^{\text{ss}}(\mathbf{r}_1,...,\mathbf{r}_N)}\right]
\nonumber \\ &=& 
- v_0 \sum_i 
\left(\mathbf{e}_i - \left<\mathbf{e}_i\right>_{\text{lss}}\right)
P_N^{\text{ss}}(\mathbf{r}_1,\mathbf{f}_1,...,\mathbf{r}_N,\mathbf{f}_N)
\cdot
\left[\boldsymbol{\nabla}_i \frac{P_N(\mathbf{r}_1,...,\mathbf{r}_N;z)}
{P_N^{\text{ss}}(\mathbf{r}_1,...,\mathbf{r}_N)}\right].
\end{eqnarray}
Similarly, one can show that
\begin{eqnarray}\label{left}
\mathcal{P}_{\text{lss}} \Omega \mathcal{Q}_{\text{lss}} ...  =  
- \frac{P_N^{\text{ss}}(\mathbf{r}_1,\theta_1,...,\mathbf{r}_N,\theta_N)}
{P_N^{\text{ss}}(\mathbf{r}_1,...,\mathbf{r}_N)} 
v_0 \sum_i \boldsymbol{\nabla}_i \cdot
\int  d\theta_1 ... d\theta_N
\left(\mathbf{e}_i - \left<\mathbf{e}_i\right>_{\text{lss}}\right) ... .
\end{eqnarray}
\end{widetext}
To proceed, we will need to deal with the projected evolution operator 
$\mathcal{Q}_{\text{lss}}\Omega\mathcal{Q}_{\text{lss}}$ in Eq. (\ref{proj1}). 
This operator 
describes evolution in the space orthogonal to the local steady-state space. The 
simplest possible approximation is to
assume that this evolution is entirely due to the free relaxation of
the self-propulsions. In this case 
$\mathcal{Q}_{\text{lss}}\Omega\mathcal{Q}_{\text{lss}}$ is approximated as 
follows
\begin{equation}\label{appQOQ}
\mathcal{Q}_{\text{lss}}\Omega\mathcal{Q}_{\text{lss}} \approx
D_r \sum_{i=1}^{N} \partial_{\theta_i}^2 .
\end{equation}
Approximation (\ref{appQOQ}) is equivalent to assuming that the relaxation in the
space orthogonal to the local steady-state space is the same in non-interacting 
and interacting systems. In particular, the approximation (\ref{appQOQ}) 
neglects the influence of the correlations between self-propulsions and 
positions on the evolution in the space orthogonal to the local steady-state space.

Combining approximation (\ref{appQOQ}) with Eqs. (\ref{right}-\ref{left}) 
we get the following approximate equality
\begin{widetext}
\begin{eqnarray}\label{proj1secondterm0}
&&
\mathcal{P}_{\text{lss}} \Omega \mathcal{Q}_{\text{lss}}
\left(z-\mathcal{Q}_{\text{lss}}\Omega\mathcal{Q}_{\text{lss}}\right)^{-1} 
\mathcal{Q}_{\text{lss}}\Omega \mathcal{P}_{\text{lss}} 
P_N(\mathbf{r}_1,\theta_1,...,\mathbf{r}_N,\theta_N;z)
\nonumber \\ && \approx
\frac{P_N^{\text{ss}}(\mathbf{r}_1,\theta_1,...,\mathbf{r}_N,\theta_N)}
{P_N^{\text{ss}}(\mathbf{r}_1,...,\mathbf{r}_N)} 
v_0^{2} \sum_i \boldsymbol{\nabla}_i \cdot
\int  d\theta_1 ... d\theta_N
\left(\mathbf{e}_i - \left<\mathbf{e}_i\right>_{\text{lss}}\right)
\nonumber \\ && \times
\left[z - D_r\sum_{j=1}^{N} \partial_{\theta_i}^2 
\right]^{-1}
\sum_l
\left(\mathbf{e}_l - \left<\mathbf{e}_l\right>_{\text{lss}}\right)
P_N^{\text{ss}}(\mathbf{r}_1,\theta_1,...,\mathbf{r}_N,\theta_N)
\cdot
\left[\boldsymbol{\nabla}_l \frac{P_N(\mathbf{r}_1,...,\mathbf{r}_N;z)}
{P_N^{\text{ss}}(\mathbf{r}_1,...,\mathbf{r}_N)}\right].
\end{eqnarray}
Now, we expand $\left[z - D_r\sum_{j=1}^{N} \partial_{\theta_i}^2 
\right]^{-1}$ and integrate by parts. In this way we get the following expression
for the second term inside the brackets on right-hand-side of Eq. (\ref{proj1}), 
\begin{eqnarray}\label{proj1secondterm}
\lefteqn{
\mathcal{P}_{\text{lss}} \Omega \mathcal{Q}_{\text{lss}}
\left(z-\mathcal{Q}_{\text{lss}}\Omega\mathcal{Q}_{\text{lss}}\right)^{-1} 
\mathcal{Q}_{\text{lss}}\Omega \mathcal{P}_{\text{lss}} 
P_N(\mathbf{r}_1,\theta_1,...,\mathbf{r}_N,\theta_N;z)}
\nonumber \\ & \approx & 
\frac{P_N^{\text{ss}}(\mathbf{r}_1,\theta_1,...,\mathbf{r}_N,\theta_N)}
{P_N^{\text{ss}}(\mathbf{r}_1,...,\mathbf{r}_N)}
v_0^{-2} \sum_{i,j} 
\boldsymbol{\nabla}_i \cdot \left(z+D_r\right)^{-1}
\left(\left<\mathbf{e}_i \mathbf{e}_j\right>_{\text{lss}} 
- \left<\mathbf{e}_i\right>_{\text{lss}} \left<\mathbf{e}_j\right>_{\text{lss}}\right)
\nonumber \\ && 
\cdot
\left\{\boldsymbol{\nabla}_j - \left[
\boldsymbol{\nabla}_j \ln P_N^{\text{ss}}(\mathbf{r}_1,...,\mathbf{r}_N) \right]\right\}
P_N(\mathbf{r}_1,...,\mathbf{r}_N;z).
\end{eqnarray}
Combining the right-hand-sides of Eqs. (\ref{proj1firstterm2}) and
(\ref{proj1secondterm}) allows us to identify the effective evolution operator
$\Omega^{\text{eff}}(z)$ governing the time-dependence of the distribution
of particles' positions,
\begin{eqnarray}\label{Omegaeff}
\Omega^{\text{eff}}(z) = \sum_{i,j} \boldsymbol{\nabla}_i \cdot 
\left[D_t \delta_{ij} + v_0^2  
\left(z+D_r\right)^{-1}
\left(\left<\mathbf{e}_i \mathbf{e}_j\right>_{\text{lss}} 
- \left<\mathbf{e}_i\right>_{\text{lss}} \left<\mathbf{e}_j
\right>_{\text{lss}}\right)\right]
\cdot
\left\{\boldsymbol{\nabla}_j- \left[
\boldsymbol{\nabla}_j \ln P_N^{\text{ss}}(\mathbf{r}_1,...,\mathbf{r}_N) \right]\right\}.
\end{eqnarray} 
\end{widetext}
The dependence of the effective evolution operator $\Omega^{\text{eff}}(z)$
on $z$ allows us to retain, albeit in an approximate way, the finite relaxation
time of the self-propulsions. This is in contrast with approximate approaches
of Farage and Brader \cite{Farage} and of Feng and Hou \cite{Feng2017}. 
In particular, the evolution operator corresponding to the former approach can 
be recovered from Eq. (\ref{Omegaeff}) by taking the $z\to 0$ limit,
\textit{i.e.} by neglecting the finite relaxation time of the self-propulsions,
and at the same time neglecting local steady-state correlations between the 
self-propulsions, $\left<\mathbf{e}_i \mathbf{e}_j\right>_{\text{lss}} 
- \left<\mathbf{e}_i\right>_{\text{lss}} \left<\mathbf{e}_j
\right>_{\text{lss}} \rightarrow \frac{1}{2}\delta_{ij} \mathbf{I}$, 
where $\mathbf{I}$ is the unit tensor. 

The above described approximations become exact in the limit of fast evolution 
of the self-propulsions, $D_r\to\infty$, $v_0\to\infty$, $v_0^2/D_r=\text{ const}$.
In this case, the ABP system becomes equivalent to a thermal system
at temperature $T + v_0^2/(2 D_r\xi_0)$. 

The finite time scale relaxation of the self-propulsions is retained in
Eq. (\ref{Omegaeff}) in the simplest possible way. One could try 
using a more sophisticated approximation in place of Eq. (\ref{appQOQ}),
for example by following the procedure described  in Appendix A of Ref. 
\cite{Szamel2016}. However, the
resulting expressions would rapidly become rather difficult to handle.
We note that, in principle, within the theory of Liluashvili \textit{et al.}
\cite{Voigtmann2017} the relaxation of self-propulsions is handled much more 
accurately \cite{commentlowd}.
However, additional technical approximations used in Ref. \cite{Voigtmann2017} make 
their description of the relaxation of self-propulsions similar to ours. 
Specifically, both in the theory of Ref. \cite{Voigtmann2017} and in our approach
the matrix representing the self-propulsion part of the evolution operator is 
effectively approximated by its lowest order non-trivial elements.


The effective evolution operator $\Omega^{\text{eff}}(z)$ allows us to re-write
the Laplace transform of the intermediate scattering function,
\begin{eqnarray}\label{Fqz}
\lefteqn{\mathcal{LT}\left[F(q;t)\right](z) \equiv 
F(q;z)} \nonumber \\ &=& 
N^{-1}\left<n(\mathbf{q})\left(z-\Omega\right)^{-1} n(-\mathbf{q})
\right>  \nonumber \\ &\approx & N^{-1}\left<n(\mathbf{q}) 
\left(z-\Omega^{\text{eff}}(z)\right)^{-1} 
n(-\mathbf{q})\right>_\mathbf{r}.
\end{eqnarray}
Here $\left< ... \right>_\mathbf{r}$ denotes averaging over the 
steady-state distribution of particles' positions. Eq. (\ref{Fqz})
will be the starting point for the standard projection operator
derivation of the memory function representation for $F(q;z)$ 
in subsection \ref{subsec:memfun}.

\subsection{Short-time dynamics and the importance of the correlations 
of particles' currents}

Before turning to the memory function representation, which focuses on the
long-time dynamics, we briefly examine the short-time dynamics of 
the intermediate scattering function and identify the contribution of the
correlations of the particles' currents. 

To evaluate the short-time behavior of the intermediate scattering
function we expand expression (\ref{Fqt}) in powers of $t$,
\begin{eqnarray}\label{Fqtst1}
F(q;t) &=&  \frac{1}{N}
\left<n(\mathbf{q}) n(-\mathbf{q})\right>
+ \frac{t}{N}
\left<n(\mathbf{q}) \Omega  n(-\mathbf{q})\right>
\nonumber \\ && + \frac{t^2}{2N}
\left<n(\mathbf{q}) \Omega^2 n(-\mathbf{q})\right> + ...
\end{eqnarray}
The first term at the right-hand-side of Eq. (\ref{Fqtst1}) is the steady
state static structure factor,
\begin{eqnarray}\label{Sq2}
S(q) = \frac{1}{N}
\left<n(\mathbf{q}) n(-\mathbf{q})\right>,
\end{eqnarray}
the second term, using the assumption of vanishing currents, gives
\begin{eqnarray}\label{Fqtst2}
&& \frac{t}{N}
\left<n(\mathbf{q}) \Omega  n(-\mathbf{q})\right> = - D_t t q^2
\end{eqnarray}
and the third term, also under the assumption of vanishing currents, reads
\begin{eqnarray}\label{Fqtst3}
&& \frac{t^2}{2N}
\left<n(\mathbf{q}) \Omega^2  n(-\mathbf{q})\right> = 
\frac{D_t^2 t^2 q^2}{2N} 
\nonumber \\ \nonumber && \times
\left<\left|\sum_i \left\{-iq + \left[\hat{\mathbf{q}}\cdot
\boldsymbol{\nabla}_j \ln P_N^{\text{ss}}(\mathbf{r}_1,...,\mathbf{r}_N) \right] \right\}
e^{-i\mathbf{q}\cdot\mathbf{r}_i} \right|^2\right>
\\ && - \frac{v_0^2 t^2 q^2}{2N} 
\left<\left|\sum_i \hat{\mathbf{q}}\cdot
\left(\mathbf{e}_i - \left<\mathbf{e}_i\right>_{\text{lss}} \right)
e^{-i\mathbf{q}\cdot\mathbf{r}_i} \right|^2\right>, 
\end{eqnarray}
where $\hat{\mathbf{q}}$ is a unit vector, $\hat{\mathbf{q}}=\mathbf{q}/q$. The 
first term in Eq. (\ref{Fqtst3}) originates from thermal fluctuations while the
second is the lowest order in time contribution of the self-propulsion.

Eq. (\ref{Fqtst3}) shows that at short times the self-propulsion always
speeds up the relaxation. The contribution of the self-propulsion
can be expressed in terms of the correlation function of particles 
currents,
\begin{eqnarray}\label{omegap}
&& \!\! \omega_{\parallel}(q) = \\ \nonumber &&
\frac{v_0^2}{N}\hat{\mathbf{q}}\cdot
\left< \sum_{i,j}\left(\mathbf{e}_i - \left<\mathbf{e}_i\right>_{\text{lss}} \right)
\left(\mathbf{e}_j - \left<\mathbf{e}_j\right>_{\text{lss}} \right)
e^{-i\mathbf{q}\cdot\left(\mathbf{r}_i-\mathbf{r}_j\right)}\right>
\cdot\hat{\mathbf{q}}.
\end{eqnarray}
The interpretation of function $\omega_{\parallel}(q)$ comes from the fact that
\begin{eqnarray}\label{current}
\lefteqn{ v_0\left(\mathbf{e}_i - \left<\mathbf{e}_i\right>_{\text{lss}}\right)
\equiv } 
\nonumber \\ && 
\xi_0^{-1} \mathbf{F}_i 
- D_t \boldsymbol{\nabla}_i \ln P_N^{\text{ss}}(\mathbf{r}_1,...,\mathbf{r}_N) 
+ v_0 \mathbf{e}_i
\end{eqnarray}
can be interpreted as a current of particle $i$ averaged over thermal noise.

The same result for the short-time dynamics are obtained if 
one starts from expression (\ref{Fqz}) for the intermediate scattering function.
In this case it is easiest to proceed in the Laplace space,
\begin{eqnarray}\label{Fqzexp}
F(q;z) &=& \frac{1}{z} F(q;t=0) + \frac{1}{z^2} F'(q;t=0)
\nonumber \\ && 
+  \frac{1}{z^3} F''(q;t=0)
+ ... . 
\end{eqnarray}
Using the effective evolution operator (\ref{Omegaeff}) in Eq. (\ref{Fqz})
we obtain the following expression for the first derivative,
\begin{eqnarray}\label{Fqz1}
&& F'(q;t=0) =
\frac{1}{N}\left<n(\mathbf{q})\Omega_t
n(-\mathbf{q})\right>
= -D_t q^2
\end{eqnarray}
where $\Omega_t$ is the translational diffusion part of the effective evolution
operator,
\begin{eqnarray}\label{Omegat}
\Omega_t=
D_t\sum_{i} \boldsymbol{\nabla}_i \cdot
\left\{\boldsymbol{\nabla}_i- \left[
\boldsymbol{\nabla}_i \ln P_N^{\text{ss}}(\mathbf{r}_1,...,\mathbf{r}_N) \right]\right\}.
\end{eqnarray}
Furthermore, for the second derivative we get
\begin{eqnarray}\label{Fqz2}
&& F''(q;t=0) = -\frac{1}{N}\left<n(\mathbf{q})\Omega_t^2
n(-\mathbf{q})\right> 
\nonumber \\ && 
+ \frac{v_0^2}{N}\left<n(\mathbf{q})\sum_{i,j} \boldsymbol{\nabla}_i \cdot 
\left(\left<\mathbf{e}_i \mathbf{e}_j\right>_{\text{lss}} 
- \left<\mathbf{e}_i\right>_{\text{lss}} \left<\mathbf{e}_j
\right>_{\text{lss}}\right)
\right . \nonumber \\ && \left. 
\cdot
\left\{\boldsymbol{\nabla}_j- \left[
\boldsymbol{\nabla}_j \ln P_N^{\text{ss}}(\mathbf{r}_1,...,\mathbf{r}_N) \right]\right\}
n(-\mathbf{q})\right>
\end{eqnarray}
which can be shown to coincide with the second derivative obtained from 
the third term of the Taylor expansion, Eq. (\ref{Fqtst3}). 

\subsection{Memory function representation}\label{subsec:memfun}

In this subsection we rewrite the formal expression (\ref{Fqz}) for the intermediate
scattering function in terms of the so-called frequency matrix
and irreducible memory matrix. The latter quantity contains all the unknown
non-trivial dynamic information about the system. The resulting expression for the
density correlation function in terms of the frequency matrix and the memory matrix 
is known as the memory function representation. 
Some formal manipulations in this subsection are the same as in Sec. VI of 
Ref. \cite{Szamel2016}. They are included here for completeness. 

To derive the memory function representation of $F(q;z)$ we use the projection
operator approach \cite{Goetzebook,CHess,SL}. We define a projection operator 
on the microscopic density
\begin{equation}\label{Pn}
\mathcal{P}_n = ... \left. n(-\mathbf{q})\right>_\mathbf{r}
\left<n(\mathbf{q})n(-\mathbf{q})\right>_\mathbf{r}^{-1}
\left< n(\mathbf{q}) ... \right. .
\end{equation}
We emphasize that projection operator $\mathcal{P}_n$ is defined in terms
of the steady-state distribution, unlike in the approaches of 
Farage and Brader \cite{Farage} and of Liluashvili \textit{et al.} \cite{Voigtmann2017}.
Next, we use the identity
\begin{eqnarray}\label{identity}
\frac{1}{z-\Omega^{\text{eff}}(z)} &=& \frac{1}{z - \Omega^{\text{eff}}(z)\mathcal{Q}_n} 
\\ \nonumber && + 
\frac{1}{z - \Omega^{\text{eff}}(z)\mathcal{Q}_n}\Omega^{\text{eff}}(z)\mathcal{P}_n 
\frac{1}{z-\Omega^{\text{eff}}(z)},
\end{eqnarray} 
where $\mathcal{Q}_n$ is the projection on the space orthogonal to that 
spanned by the microscopic density, 
\begin{equation}
\mathcal{Q}_n = \mathcal{I} - \mathcal{P}_n,
\end{equation}
to rewrite the Laplace transform of the time derivative of $N F(q;t)$ in the following 
way
\begin{widetext}
\begin{eqnarray}\label{timeder}
&& \mathcal{LT}[\partial_t N F(q;t)](z)= 
\left<n(\mathbf{q}) \Omega^{\text{eff}}(z) 
\frac{1}{z-\Omega^{\text{eff}}(z)} n(-\mathbf{q}) \right>_\mathbf{r} 
= \left<n(\mathbf{q}) \Omega^{\text{eff}}(z) \mathcal{P}_n 
\frac{1}{z-\Omega^{\text{eff}}(z)} n(-\mathbf{q}) \right>_\mathbf{r}
\nonumber \\ && + \left<n(\mathbf{q}) \Omega^{\text{eff}}(z) \mathcal{Q}_n 
\frac{1}{z-\Omega^{\text{eff}}(z)} n(-\mathbf{q}) \right>_\mathbf{r} 
=
\left<n(\mathbf{q}) \Omega^{\text{eff}}(z) n(-\mathbf{q})\right>_\mathbf{r}
\left<n(\mathbf{q})n(-\mathbf{q})\right>_\mathbf{r}^{-1}
\left< n(\mathbf{q}) \frac{1}{z-\Omega^{\text{eff}}(z)} n(-\mathbf{q}) 
\right>_\mathbf{r} 
\nonumber \\ && + 
\left<n(\mathbf{q}) \Omega^{\text{eff}}(z) \mathcal{Q}_n 
\frac{1}{z - \mathcal{Q}_n\Omega^{\text{eff}}(z)\mathcal{Q}_n} 
\mathcal{Q}_n \Omega^{\text{eff}}(z) n(-\mathbf{q})\right>_\mathbf{r}
\left<n(\mathbf{q})n(-\mathbf{q})\right>_\mathbf{r}^{-1}
\left< n(\mathbf{q})\frac{1}{z-\Omega^{\text{eff}}(z)}
n(-\mathbf{q}) \right>_\mathbf{r}.
\end{eqnarray}
\end{widetext}
The important part of the first term on the right-hand-side of the last equality sign in 
Eq. (\ref{timeder}) is the matrix element of the effective evolution operator, 
$\left<n(\mathbf{q}) \Omega^{\text{eff}}(z) n(-\mathbf{q})\right>_\mathbf{r}$, 
which can be expressed in terms of the frequency matrix $\mathcal{H}(q;z)$,
\begin{eqnarray}\label{freqmat1}
\left<n(\mathbf{q}) \Omega^{\text{eff}}(z) n(-\mathbf{q})\right>_\mathbf{r}
= -q^2 N \mathcal{H}(q;z).
\end{eqnarray}
The frequency matrix is given by the following expression
\begin{eqnarray}\label{freqmat2}
&& \mathcal{H}(q;z) =  D_t + \frac{\omega_\parallel(q)}{z+D_r}.
\end{eqnarray}

The important part of the second term at the right-hand-side of the last equality 
sign in Eq. (\ref{timeder}) can be expressed in terms of reducible \cite{CHess,Kawasaki} 
memory matrix $\mathcal{M}(q;z)$,
\begin{eqnarray}\label{memfunred1}
&& \left<n(\mathbf{q}) \Omega^{\text{eff}}(z) \mathcal{Q}_n 
\frac{1}{z - \mathcal{Q}_n\Omega^{\text{eff}}(z)\mathcal{Q}_n} 
\mathcal{Q}_n \Omega^{\text{eff}}(z) n(-\mathbf{q})\right>_{\mathbf{r}} =
\nonumber \\ && q^2 N \mathcal{M}(q;z).
\end{eqnarray}
Explicitly, the memory matrix is given by the following expression
\begin{widetext}
\begin{eqnarray}\label{memfunred2}
&&  \mathcal{M}(q;z) = 
N^{-1}
\hat{\mathbf{q}}\cdot\left< \sum_{i,j} e^{-i\mathbf{q}\cdot\mathbf{r}_i}
\left[D_t \delta_{ij} + \frac{v_0^2}{z+D_r}\left(
\left<\mathbf{e}_i \mathbf{e}_j\right>_{\text{lss}} 
- \left<\mathbf{e}_i\right>_{\text{lss}} 
\left<\mathbf{e}_j\right>_{\text{lss}}\right)\right]\cdot
\left\{-\boldsymbol{\nabla}_j + \left[
\boldsymbol{\nabla}_j \ln P_N^{\text{ss}}(\mathbf{r}_1,...,\mathbf{r}_N) \right]\right\}
\mathcal{Q}_n
\right. \nonumber \\ && \times \left.
\frac{1}{z - \mathcal{Q}_n\Omega^{\text{eff}}(z)\mathcal{Q}_n}
\mathcal{Q}_n \sum_{l,m} 
\boldsymbol{\nabla}_l \cdot \left[D_t \delta_{lm} + 
\frac{v_0^2}{z+D_r}\left(\left<\mathbf{f}_l \mathbf{f}_m\right>_{\text{lss}} 
- \left<\mathbf{f}_l\right>_{\text{lss}} \left<\mathbf{f}_m\right>_{\text{lss}}\right)
\right]e^{i\mathbf{q}\cdot\mathbf{r}_m} \right>_{\mathbf{r}}\cdot\hat{\mathbf{q}}
\end{eqnarray}
\end{widetext}

We can now rewrite the Laplace transform of the intermediate
scattering function in terms of the frequency and memory matrix,
\begin{eqnarray}\label{fqzmfred}
F(q;z) = \frac{S(q)}{z+q^2\left(\mathcal{H}(q;z) - \mathcal{M}(q;z)\right)/S(q)} 
\end{eqnarray}
where $S(q)$ is the steady-state structure factor,
\begin{eqnarray}\label{Sq}
S(q)= \left<n(\mathbf{q})n(-\mathbf{q})\right>_\mathbf{r} 
\equiv \left<n(\mathbf{q})n(-\mathbf{q})\right>.
\end{eqnarray}
The second equality sign in Eq.~(\ref{Sq}) follows from the fact that for 
self-propulsion-independent quantities averaging over particles' positions 
is equivalent to averaging over the full steady-state distribution of
positions and self-propulsions. 

Next, following Cichocki and Hess \cite{CHess} and Kawasaki \cite{Kawasaki} we
introduce an irreducible memory matrix. 
First, we define the irreducible evolution operator $\Omega^{\mathrm{irr}}(z)$,
\begin{equation}\label{Omegairr1}
\Omega^{\text{irr}}(z) = \mathcal{Q}_n\Omega^{\text{eff}}(z)\mathcal{Q}_n -
\delta \Omega^{\text{irr}}(z)
\end{equation}
where the subtraction term  $\delta \Omega^{\text{irr}}(z)$ reads 
\begin{widetext}
\begin{eqnarray}\label{Omegairr2}
\delta \Omega^{\text{irr}}(z) &=& \left. \mathcal{Q}_n \sum_{l,m} 
\boldsymbol{\nabla}_l \cdot \left[D_t \delta_{lm} + 
\frac{v_0^2}{z+D_r}\left(\left<\mathbf{f}_l \mathbf{f}_m\right>_{\text{lss}} 
- \left<\mathbf{f}_l\right>_{\text{lss}} \left<\mathbf{f}_m\right>_{\text{lss}}\right)
\right]e^{i\mathbf{q}\cdot\mathbf{r}_m} \right>_{\mathbf{r}}\cdot\hat{\mathbf{q}}
\left(\mathcal{H}(q;z)\right)^{-1}
\nonumber \\ && \hat{\mathbf{q}}\cdot\left< \sum_{i,j} e^{-i\mathbf{q}\cdot\mathbf{r}_i}
\left[D_t \delta_{ij} + \frac{v_0^2}{z+D_r}\left(
\left<\mathbf{e}_i \mathbf{e}_j\right>_{\text{lss}} 
- \left<\mathbf{e}_i\right>_{\text{lss}} 
\left<\mathbf{e}_j\right>_{\text{lss}}\right)\right]\cdot
\left\{-\boldsymbol{\nabla}_j + \left[
\boldsymbol{\nabla}_j \ln P_N^{\text{ss}}(\mathbf{r}_1,...,\mathbf{r}_N) \right]\right\}
\mathcal{Q}_n
\right. .
\end{eqnarray}
Next, we define the irreducible memory matrix 
$\mathcal{M}^{\text{irr}}(q;z)$, which is given by the expression analogous to 
Eq. (\ref{memfunred2}) but with the projected evolution operator
$\mathcal{Q}_n\Omega^{\text{eff}}(z)\mathcal{Q}_n$ replaced by 
irreducible evolution  operator $\Omega^{\text{irr}}(z)$,
\begin{eqnarray}\label{memfunirr1}
&&  \mathcal{M}^{\text{irr}}(q;z) = 
N^{-1}
\hat{\mathbf{q}}\cdot\left< \sum_{i,j} e^{-i\mathbf{q}\cdot\mathbf{r}_i}
\left[D_t \delta_{ij} + \frac{v_0^2}{z+D_r}\left(
\left<\mathbf{e}_i \mathbf{e}_j\right>_{\text{lss}} 
- \left<\mathbf{e}_i\right>_{\text{lss}} 
\left<\mathbf{e}_j\right>_{\text{lss}}\right)\right]\cdot
\left\{-\boldsymbol{\nabla}_j + \left[
\boldsymbol{\nabla}_j \ln P_N^{\text{ss}}(\mathbf{r}_1,...,\mathbf{r}_N) \right]\right\}
\mathcal{Q}_n
\right. \nonumber \\ && \times \left.
\frac{1}{z - \Omega^{\text{irr}}(z)}
\mathcal{Q}_n \sum_{l,m} 
\boldsymbol{\nabla}_l \cdot \left[D_t \delta_{lm} + 
\frac{v_0^2}{z+D_r}\left(\left<\mathbf{f}_l \mathbf{f}_m\right>_{\text{lss}} 
- \left<\mathbf{f}_l\right>_{\text{lss}} \left<\mathbf{f}_m\right>_{\text{lss}}\right)
\right]e^{i\mathbf{q}\cdot\mathbf{r}_m} \right>_{\mathbf{r}}\cdot\hat{\mathbf{q}}.
\end{eqnarray}
\end{widetext}
Finally, we use an identity similar to Eq. (\ref{identity}),
\begin{eqnarray}\label{identity2}
\lefteqn{\frac{1}{z-\mathcal{Q}_n\Omega^{\text{eff}}(z)\mathcal{Q}_n} 
= \frac{1}{z - \Omega^{\text{irr}}(z)} }
\\ \nonumber && + 
\frac{1}{z - \Omega^{\text{irr}}(z)}\delta\Omega^{\text{eff}}(z) 
\frac{1}{z-\mathcal{Q}_n\Omega^{\text{eff}}(z)\mathcal{Q}_n},
\end{eqnarray} 
and we derive the following relation between $\mathcal{M}(q;z)$ and 
$\mathcal{M}^{\text{irr}}(q;z)$,
\begin{equation}\label{redirr}
\mathcal{M}(q;z)=\mathcal{M}^{\text{irr}}(q;z)-\mathcal{M}^{\text{irr}}(q;z)
\mathcal{H}^{-1}(q;z)\mathcal{M}(q;z).
\end{equation}

Combining Eqs. (\ref{fqzmfred}) and (\ref{redirr}) we arrive at the following
representation of the intermediate scattering function in terms of the
irreducible memory matrix,
\begin{eqnarray}\label{fqzmfirr}
F(q;z) = \frac{S(q)}{z+\frac{q^2\mathcal{H}(q;z)/S(q)}
{1+\mathcal{M}^{\text{irr}}(q;z)/\mathcal{H}(q;z)}}.
\end{eqnarray}

Eq. (\ref{fqzmfirr}) constitutes the memory function representation for the 
intermediate scattering function of the ABP system. It has the same structure
as the memory function representation for the intermediate scattering function 
of the AOUP model. The difference between the two models is buried in the 
expressions for the frequency and irreducible memory matrices. 

\subsection{Mode-coupling approximation for the irreducible memory matrix}

Here, we derive an explicit approximate expression for the irreducible
memory matrix using a factorization approximation of the type used in the
mode-coupling theory of the glass transition. To this end we follow the steps 
of the derivation of the mode-coupling theory for systems
evolving with Brownian dynamics \cite{SL}. The derivation consists of three steps. 

First, we project onto the subspace of density pairs,
\begin{widetext}
\begin{eqnarray}\label{projdenpairs}
\lefteqn{\hat{\mathbf{q}}\cdot\left< \sum_{i,j} e^{-i\mathbf{q}\cdot\mathbf{r}_i}
\left[D_t \delta_{ij} + \frac{v_0^2}{z+D_r}\left(
\left<\mathbf{e}_i \mathbf{e}_j\right>_{\text{lss}} 
- \left<\mathbf{e}_i\right>_{\text{lss}} 
\left<\mathbf{e}_j\right>_{\text{lss}}\right)\right]\cdot
\left\{-\boldsymbol{\nabla}_j + \left[
\boldsymbol{\nabla}_j \ln P_N^{\text{ss}}(\mathbf{r}_1,...,\mathbf{r}_N) \right]\right\}
\mathcal{Q}_n \right. }
\nonumber \\ \nonumber && \!\!\!\!\!\!\!\! \approx \sum_{\mathbf{q}_1,...,\mathbf{q}_4}
\hat{\mathbf{q}}\cdot\left< \sum_{i,j} e^{-i\mathbf{q}\cdot\mathbf{r}_i}
\left[D_t \delta_{ij} + \frac{v_0^2}{z+D_r}\left(
\left<\mathbf{e}_i \mathbf{e}_j\right>_{\text{lss}} 
- \left<\mathbf{e}_i\right>_{\text{lss}} 
\left<\mathbf{e}_j\right>_{\text{lss}}\right)\right]\cdot
\left\{-\boldsymbol{\nabla}_j + \left[
\boldsymbol{\nabla}_j \ln P_N^{\text{ss}}(\mathbf{r}_1,...,\mathbf{r}_N) \right]\right\}
\mathcal{Q}_n n_2(-\mathbf{q}_1,-\mathbf{q}_2)\right>_{\mathbf{r}} 
\\ && \!\!\!\!\!\!\!\!  \times 
\left[\left<\mathcal{Q}_n n_2(\mathbf{q}_1,\mathbf{q}_2) \mathcal{Q}_n 
n_2(-\mathbf{q}_3,-\mathbf{q}_4)
\right>_{\mathbf{r}}\right]^{-1}\left< \mathcal{Q}_n n_2(\mathbf{q}_3,\mathbf{q}_4)
\right. .
\end{eqnarray}
\end{widetext}
Here $n_2(\mathbf{q}_1,\mathbf{q}_2)$ is the Fourier transform of the microscopic 
two-particle density,
\begin{equation}\label{n2def}
n_2(\mathbf{q}_1,\mathbf{q}_2) = \sum_{l,m} 
e^{-i\mathbf{q}_1\cdot\mathbf{r}_l-i\mathbf{q}_2\cdot\mathbf{r}_m},
\end{equation}
and $\left[\left<\mathcal{Q}_n n_2(\mathbf{q}_1,\mathbf{q}_2) \mathcal{Q}_n 
n_2(-\mathbf{q}_3,-\mathbf{q}_4)
\right>_{\mathbf{r}}\right]^{-1}$ is the inverse of the correlation matrix of 
microscopic pair densities. 

Second, we factorize averages resulting from substituting projection
(\ref{projdenpairs}) into the expression for the memory function and
\textit{at the same time} replace the irreducible operator $\Omega^{\text{irr}}(z)$
by effective evolution operator $\Omega^{\text{eff}}(z)$. We should emphasize that 
this factorization has to be done in the time domain,
\begin{widetext}
\begin{eqnarray}\label{fac4to2}
\lefteqn{\mathcal{LT}^{-1}\left[\left< \mathcal{Q}_n n_2(\mathbf{q}_1,\mathbf{q}_2) 
\left(z-\Omega^{\text{irr}}(z)\right)^{-1} \mathcal{Q}_n 
n_2(-\mathbf{q}_3,-\mathbf{q}_4)\right>_{\mathbf{r}}\right] \approx}
\\ \nonumber && 
\mathcal{LT}^{-1}\left[\left< n(\mathbf{q}_1) 
\left(z-\Omega^{\text{eff}}(z)\right)^{-1} n(-\mathbf{q}_3)\right>_{\mathbf{r}}\right]
\mathcal{LT}^{-1}\left[\left< n(\mathbf{q}_2) 
\left(z-\Omega^{\text{eff}}(z)\right)^{-1} n(-\mathbf{q}_4)\right>_{\mathbf{r}}\right]
+ \left\{ 3 \leftrightarrow 4 \right\}.
\end{eqnarray}
\end{widetext}
Here $\mathcal{LT}^{-1}$ denotes the inverse Laplace transform and
$\left\{ 3 \leftrightarrow 4 \right\}$ means the preceding expression with
labels $3$ and $4$ interchanged. Consistently with
Eq. (\ref{fac4to2}) we also factorize the steady-state correlation matrix
of microscopic pair densities and for its inverse we get
\begin{eqnarray}\label{fac4to2ss}
&& \left[\left<\mathcal{Q}_n n_2(\mathbf{q}_1,\mathbf{q}_2) \mathcal{Q}_n 
n_2(-\mathbf{q}_3,-\mathbf{q}_4)
\right>_{\mathbf{r}}\right]^{-1} \approx
\\ \nonumber && 
\left<n(\mathbf{q}_1) n(-\mathbf{q}_3)\right>_{\mathbf{r}}^{-1}
\left<n(\mathbf{q}_2) n(-\mathbf{q}_4)\right>_{\mathbf{r}}^{-1}
+ \left\{ 3 \leftrightarrow 4 \right\}.
\end{eqnarray}

Third, we approximate the vertex functions. Due to the presence of the current
correlations in one part of the vertex, this last step is a bit more complex 
than the approximation used in the derivation of the standard mode-coupling theory 
\cite{SL}. We will explain it on the 
example of the left vertex, $\mathcal{V}_l$. 

The left vertex is given by the following formula
\begin{eqnarray}\label{leftvertexdef}
\lefteqn{ \mathcal{V}_l(\mathbf{q};\mathbf{q}_1,\mathbf{q}_2) = 
\hat{\mathbf{q}}\cdot\left< \sum_{i,j} e^{-i\mathbf{q}\cdot\mathbf{r}_i} 
\right. }
\nonumber \\ && \times \left.
\left[D_t \delta_{ij} + \frac{v_0^2}{z+D_r}\left(
\left<\mathbf{e}_i \mathbf{e}_j\right>_{\text{lss}} 
- \left<\mathbf{e}_i\right>_{\text{lss}} 
\left<\mathbf{e}_j\right>_{\text{lss}}\right)\right]\cdot
\right.\nonumber \\ && \times \left.
\left\{-\boldsymbol{\nabla}_j + \left[
\boldsymbol{\nabla}_j \ln P_N^{\text{ss}}(\mathbf{r}_1,...,\mathbf{r}_N) \right]\right\}
\mathcal{Q}_n n_2(-\mathbf{q}_1,-\mathbf{q}_2)\right>_{\mathbf{r}}
\nonumber \\ &=&
\hat{\mathbf{q}}\cdot\left< \sum_{i,j} e^{-i\mathbf{q}\cdot\mathbf{r}_i}
\right.\nonumber \\ && \times \left.
\left[D_t \delta_{ij} + \frac{v_0^2}{z+D_r}\left(
\mathbf{e}_i-\left<\mathbf{e}_i\right>_{\text{lss}}\right)
\left(\mathbf{e}_j-\left<\mathbf{e}_j\right>_{\text{lss}}\right)\right]\cdot
\right.\nonumber \\ && \times \left.
\left\{-\boldsymbol{\nabla}_j + \left[
\boldsymbol{\nabla}_j \ln P_N^{\text{ss}}(\mathbf{r}_1,...,\mathbf{r}_N) \right]\right\}
\mathcal{Q}_n n_2(-\mathbf{q}_1,-\mathbf{q}_2)\right>
\nonumber \\ &&
\end{eqnarray} 

According to Eq. (\ref{leftvertexdef}) the left vertex consists of
two terms. The first term originates from the translational diffusion part 
of the effective evolution operator. It has the same form as the vertex of the 
standard mode-coupling theory \cite{SL},
\begin{eqnarray}\label{leftvertext}
\lefteqn{\mathcal{V}_l^t(\mathbf{q};\mathbf{q}_1,\mathbf{q}_2) =}
\nonumber \\ &&
D_t \hat{\mathbf{q}}\cdot\left< \sum_{i} e^{-i\mathbf{q}\cdot\mathbf{r}_i}
\left\{-\boldsymbol{\nabla}_j + \left[
\boldsymbol{\nabla}_j \ln P_N^{\text{ss}}(\mathbf{r}_1,...,\mathbf{r}_N) \right]\right\}
\right.\nonumber \\ && \times \left.
\mathcal{Q}_n n_2(-\mathbf{q}_1,-\mathbf{q}_2)\right>
\nonumber \\ &=& 
i D_t N \rho S(q_1)S(q_2)
\left[ \hat{\mathbf{q}}\cdot\mathbf{q}_1 c(q_1) +
\hat{\mathbf{q}}\cdot\mathbf{q}_2 c(q_2)\right],
\end{eqnarray}
where $c(q)$ is the direct correlation function \cite{HansenMcDonald}
\begin{equation}\label{oldc}
\rho c(q) = 1-1/S(q).
\end{equation}
We note that in the derivation of formula (\ref{leftvertex}) a convolution
approximation is used, which is equivalent to neglecting the 
three-particle direct correlation function. The same approximation is used
in the derivation of the standard mode-coupling theory \cite{SL}.

The second term originates from the activity-related part 
of the effective evolution operator. It is proportional to
function $\mathcal{V}_l^a$, which is defined as follows,
\begin{eqnarray}\label{leftvertexa1}
\lefteqn{\mathcal{V}_l^a(\mathbf{q};\mathbf{q}_1,\mathbf{q}_2) =}
\nonumber \\ && v_0^2 D_r^{-1}
\hat{\mathbf{q}}\cdot\left< \sum_{i,j} e^{-i\mathbf{q}\cdot\mathbf{r}_i}
\left(
\mathbf{e}_i-\left<\mathbf{e}_i\right>_{\text{lss}}\right)
\left(\mathbf{e}_j-\left<\mathbf{e}_j\right>_{\text{lss}}\right)\cdot
\right.\nonumber \\ && \times \left.
\left\{-\boldsymbol{\nabla}_j + \left[
\boldsymbol{\nabla}_j \ln P_N^{\text{ss}}(\mathbf{r}_1,...,\mathbf{r}_N) \right]\right\}
\mathcal{Q}_n n_2(-\mathbf{q}_1,-\mathbf{q}_2)\right>.
\nonumber \\ &&
\end{eqnarray}
$\mathcal{V}_l^a(\mathbf{q};\mathbf{q}_1,\mathbf{q}_2)$ has the form very similar
to the form of the vertex of the mode-coupling theory for the AOUP model 
\cite{Szamel2016} and can be analyzed in the same way. The result is 
\begin{eqnarray}\label{leftvertexa2}
\lefteqn{\mathcal{V}_l^a(\mathbf{q};\mathbf{q}_1,\mathbf{q}_2) =}
\nonumber \\ &=& 
i D_r^{-1} N \rho S(q_1)S(q_2) \omega_{\parallel}(q)
\left[ \hat{\mathbf{q}}\cdot\mathbf{q}_1 \mathcal{C}(q_1) +
\hat{\mathbf{q}}\cdot\mathbf{q}_2 \mathcal{C}(q_2)\right].
\nonumber \\ &&
\end{eqnarray}
where a new function $\mathcal{C}(q)$ reads
\begin{eqnarray}\label{newc}
\rho \mathcal{C}(q) = 1-\frac{\omega_{\parallel}(q)}
{\omega_{\parallel}(\infty )S(q)}.
\end{eqnarray}
Again, we note that the derivation of formula (\ref{leftvertexa2}) involves
a generalization of the convolution approximation introduced in Ref. \cite{Szamel2016}. 

Combining Eqs. (\ref{leftvertext}) and (\ref{leftvertexa2}) we get the
following approximate expression for the left vertex,
\begin{eqnarray}\label{leftvertex}
\lefteqn{ \mathcal{V}_l(\mathbf{q};\mathbf{q}_1,\mathbf{q}_2) =
\mathcal{V}_l^t(\mathbf{q};\mathbf{q}_1,\mathbf{q}_2)+
\frac{\mathcal{V}_l^a(\mathbf{q};\mathbf{q}_1,\mathbf{q}_2)}{1+z/D_r}
}
\nonumber \\ &=&
i D_t  N\rho S(q_1)S(q_2) 
\left[ \hat{\mathbf{q}}\cdot\mathbf{q}_1 c(q_1) +
\hat{\mathbf{q}}\cdot\mathbf{q}_2 c(q_2)\right]
\nonumber \\ && +
\frac{i D_r^{-1} N\rho S(q_1)S(q_2) \omega_{\parallel}(q)
\left[ \hat{\mathbf{q}}\cdot\mathbf{q}_1 \mathcal{C}(q_1) +
\hat{\mathbf{q}}\cdot\mathbf{q}_2 \mathcal{C}(q_2)\right]}{1+z/D_r}.
\nonumber \\ &&
\end{eqnarray}

The right vertex can be analyzed in the same way resulting in 
\begin{eqnarray}\label{rightvertex}
\lefteqn{ \mathcal{V}_r(\mathbf{q};\mathbf{q}_1,\mathbf{q}_2) =
\mathcal{V}_r^t(\mathbf{q};\mathbf{q}_1,\mathbf{q}_2)+
\frac{\mathcal{V}_r^a(\mathbf{q};\mathbf{q}_1,\mathbf{q}_2)}{1+z/D_r}
}
\nonumber \\ &=&
-i D_t  N\rho S(q_1)S(q_2) 
\left[ \hat{\mathbf{q}}\cdot\mathbf{q}_1 c(q_1) +
\hat{\mathbf{q}}\cdot\mathbf{q}_2 c(q_2)\right]
\nonumber \\ && -
\frac{i D_r^{-1} N\rho S(q_1)S(q_2) \omega_{\parallel}(q)
\left[ \hat{\mathbf{q}}\cdot\mathbf{q}_1 \mathcal{C}(q_1) +
\hat{\mathbf{q}}\cdot\mathbf{q}_2 \mathcal{C}(q_2)\right]}{1+z/D_r}.
\nonumber \\ &&
\end{eqnarray}
We note that both $\mathcal{V}_l$ and $\mathcal{V}_r$ are $z-$ or time-dependent. 
The time dependence originates from the finite relaxation rate of the self-propulsion.

To simplify the formulae below, we factor out $\pm iN\rho S(q_1)S(q_2)$ from
the left and right vertex and we define the vertex function
$\mathcal{V}(\mathbf{q};\mathbf{q}_1,\mathbf{q}_2)$, which is analogous to
the standard vertex functions of mode-coupling theories, 
\begin{eqnarray}\label{vertex}
\lefteqn{ \mathcal{V}(\mathbf{q};\mathbf{q}_1,\mathbf{q}_2) =
\mathcal{V}^t(\mathbf{q};\mathbf{q}_1,\mathbf{q}_2)+
\frac{\mathcal{V}^a(\mathbf{q};\mathbf{q}_1,\mathbf{q}_2)}{1+z/D_r}
}
\nonumber \\ &=&
D_t  
\left[ \hat{\mathbf{q}}\cdot\mathbf{q}_1 c(q_1) +
\hat{\mathbf{q}}\cdot\mathbf{q}_2 c(q_2)\right]
\nonumber \\ && +
\frac{D_r^{-1}  \omega_{\parallel}(q)
\left[ \hat{\mathbf{q}}\cdot\mathbf{q}_1 \mathcal{C}(q_1) +
\hat{\mathbf{q}}\cdot\mathbf{q}_2 \mathcal{C}(q_2)\right]}{1+z/D_r}.
\nonumber \\ &&
\end{eqnarray}

Now, combining the three steps constituting the mode-coupling approximation, 
expressing the vertices in terms of $\mathcal{V}^t$
and $\mathcal{V}^a$ and taking the thermodynamic limit we arrive
at the following expression for the irreducible memory matrix,
\begin{eqnarray}\label{memfction}
\mathcal{M}^{\text{irr}}(q;t) &=&
\frac{\rho}{2} \int \frac{d\mathbf{q}_1 d\mathbf{q}_2}{(2\pi)^2}
\delta(\mathbf{q}-\mathbf{q}_1-\mathbf{q}_2) \\ \nonumber && \times
\left(\mathcal{V}^t(\mathbf{q};\mathbf{q}_1,\mathbf{q}_2)
+\mathcal{V}^a(\mathbf{q};\mathbf{q}_1,\mathbf{q}_2)D_r e^{-D_rt}\ast\right)^2
\\ \nonumber && \times F(q_1;t)F(q_2;t),
\end{eqnarray}
where $\ast$ denotes convolution in the time domain. 

Formula (\ref{memfction}) describes the standard mode-coupling dynamic feedback 
mechanism: the time-delayed internal friction 
arising due to interparticle interactions decays due to the relaxation of 
the two-particle density, which is included at the level of factorization 
approximation (\ref{fac4to2}). Through formula (\ref{memfction}), slow decay of 
the density fluctuations feeds back into slow decay of the irreducible memory matrix,
which quantifies the internal friction.

The non-equilibrium nature of the active system is included at two levels. First,
since the microscopic dynamics consists of random thermal motion and motion due to
the independently evolving self-propulsion, the vertices exhibit two contributions
corresponding to these two microscopic modes of motion. Second, the contribution
due to the self-propulsion $\mathcal{V}^a$, involves equal-time correlation function
of particles' currents, $\omega_{\parallel}(q)$. This function sets the overall
scale of the self-propulsion contribution (together with $D_r$). It also enters
into the expression for the new function $\mathcal{C}(q)$, Eq. (\ref{newc}), which
plays the role of the direct correlation function in $\mathcal{V}^a$. 

\subsection{Long-time dynamics and ergodicity-breaking transition}

To obtain the time dependence of the intermediate scattering function one needs to 
solve the combined set of Eqs. (\ref{fqzmfirr}) and (\ref{memfction}). In the 
case of AOUPs, it is possible to rewrite the analogous set of equations and to get 
a self-consistent equation of motion for the intermediate scattering function that
has the form very similar to that of the mode-coupling equation for the intermediate 
scattering function of an under-damped thermal colloidal system \cite{Szamel2016}. 
In the present case,
with microscopic dynamics due to two independent mechanisms, this does not seem 
possible. The exception is the asymptotic long-time dynamics close to the 
ergodicity-breaking transition, which is discussed below. 

Generically, equations similar to (\ref{fqzmfirr}) and (\ref{memfction}) predict
that as equal-time correlations grow (which can happen by lowering the temperature,
increasing the number density, or, in the present case, by manipulating
the self-propulsion), the memory matrix grows and leads to the slowing down of
the time evolution of the intermediate scattering function. Eventually,
the the equations predict an ergodicity breaking transition, at which the intermediate 
scattering function ceases decaying to zero. Close to the transition and at long
times (which corresponds to small Laplace variable $z$), Eqs. (\ref{fqzmfirr}) and 
(\ref{memfction}) can be approximated by a simpler set of equations. These 
asymptotic equations are usually expressed in terms of the so-called normalized
correlator, $\phi(q;t)=F(q;t)/S(q)$, and normalized irreducible memory function,
$m(q;t)$.  They have the following form, 
\begin{eqnarray}\label{phiqzasympt}
\frac{\phi(q;z)}{1-z\phi(q;z)} = m(q;z),
\end{eqnarray}
\begin{eqnarray}\label{memfctionasympt}
&& m(q;t)=
\frac{\rho S(q)}{2q^2 \left(D_t+\frac{\omega_\parallel(q)}{D_r}\right)^2} 
\int \frac{d\mathbf{q}_1 d\mathbf{q}_2}{(2\pi)^2}
\delta(\mathbf{q}-\mathbf{q}_1-\mathbf{q}_2) 
\nonumber \\ && \times
\left\{\hat{\mathbf{q}}\cdot
\left[ D_t c(q_1)+\frac{\omega_\parallel(q_1)}{D_r}\mathcal{C}(q_1)\right]
\mathbf{q}_1 
+ \left\{1\leftrightarrow 2\right\}\right\}^2
\nonumber \\ && \times S(q_1)S(q_2)\phi(q_1;t)\phi(q_2;t),
\end{eqnarray}
Eqs. (\ref{phiqzasympt}-\ref{memfctionasympt}) have the structure very
similar to that of the self-consistent equations of motion for the 
long-time dynamics near the ergodicity breaking transition described by the standard 
mode-coupling theory. This implies that all analytical results based of the
standard mode-coupling theory can be used for the present theory for the
dynamics of the ABP system. In particular, the only quantity that one needs to
calculate in order to predict mode-coupling exponents is the so-called 
exponent parameter $\lambda$ \cite{Goetzebook}. This parameter can be calculated
from the solution of the self-consistent equations for the order parameter at 
the ergodicity breaking transition.

The reduced memory function $m(q;z)$ differs from the corresponding quantity 
of the standard mode-coupling theory by the fact that the vertex involves
a weighted average of the direct correlation function and the new function
$\mathcal{C}(q)$, which involves $S(q)$ and $\omega_{\parallel}(q)$.
We emphasize that it is the non-equilibrium steady-state structure factor
that enters into Eq. (\ref{memfctionasympt}) and determines $c(q)$ and
$\mathcal{C}(q)$. However, unlike in the standard mode-coupling theory, the 
structure factor itself does not completely determine the system's dynamics,
since the memory function involves also the correlation function of particles'
currents $\omega_\parallel(q)$.

Finally, let us assume that the system, as described by Eqs. 
(\ref{phiqzasympt}-\ref{memfctionasympt}), undergoes an ergodicity 
breaking transition. At such a transition, the normalized correlator ceases decaying 
to zero. The long-time limit of the correlator, $\lim_{t\to\infty} \phi(q;t) = f(q)$,
is the order parameter of the non-ergodic state. Eqs. 
(\ref{phiqzasympt}-\ref{memfctionasympt}) lead to the following set of 
self-consistent equations for the order parameter, $f(q)$, 
\begin{eqnarray}\label{fq}
\frac{f(q)}{1-f(q)} = m(q)
\end{eqnarray}
where $m(q)$ is given by the following equation
\begin{eqnarray}\label{mq}
&& m(q) = \frac{\rho S(q)}{2q^2 \left(D_t+\frac{\omega_\parallel(q)}{D_r}\right)^2} 
\int \frac{d\mathbf{q}_1 d\mathbf{q}_2}{(2\pi)^3}
\delta(\mathbf{q}-\mathbf{q}_1-\mathbf{q}_2) 
\nonumber \\ && \times
\left\{\hat{\mathbf{q}}\cdot
\left[ D_t c(q_1)+\frac{\omega_\parallel(q_1)}{D_r}\mathcal{C}(q_1)\right]
\mathbf{q}_1 
+ \left\{1\leftrightarrow 2\right\}\right\}^2
\nonumber \\ && \times S(q_1)S(q_2)f(q_1)f(q_2).
\end{eqnarray}
Again, self-consistent equations (\ref{fq}-\ref{mq})
for the order parameter are very similar to 
the equations derived in the standard mode-coupling theory. The non-equilibrium 
character of the ABP system manifests itself in the second equation. 

We note that, unlike in the integration-through-transients theory of 
Liluashvili \textit{et al.} \cite{Voigtmann2017}, the ergodicity-breaking
transition is determined only by equal-time quantities characterizing 
the steady state of the ABP system.

Finally, we recall that in the limit of rapidly relaxing self-propulsion,
$D_r\to\infty$, $v_0\to\infty$, $v_0^2/D_r = \text{ const.}$, the ABP
system becomes equivalent to a thermal system at temperature equal to the 
the single-particle effective temperature,
$T_{\text{eff}}= T + v_0^2/\left(2 D_r\xi_0\right)$. We note that
in this limit the static structure factor becomes equal to the equilibrium
structure factor at $T_\text{eff}$. Moreover, $\mathcal{C}(q)$ becomes equal to
the direct correlation function. With these two changes, 
Eqs. (\ref{phiqzasympt}-\ref{memfctionasympt})
coincide with standard mode-coupling equations for a thermal system at temperature 
$T_\text{eff}$.

\section{Discussion}\label{sec:disc}

We presented here a theory for the steady state dynamics of dense systems of active
Brownian particles. The derivation identified a function that quantifies
correlations of steady state particles' currents, which influences both short- and 
long-time dynamics of the ABP system. 

The present theory for the dynamics of ABP systems relies upon a factorization
approximation. Thus, like our previous theory for the dynamics of AOUP systems,
it belongs to the class of mode-coupling-like theories. Our theory predicts that 
the dynamics upon approaching an ergodicity breaking transition is qualitatively 
similar to that predicted by the standard mode-coupling
theory close to the corresponding mode-coupling transition. Quantitative details,
including the location of the transition, the exponent parameter $\lambda$ and 
the mode-coupling exponents, depend on both the steady-state structure factor and 
the correlation function of steady-state currents.

In the limit of vanishing self-propulsion the theory reduces itself to the
standard mode-coupling theory for colloidal glassy dynamics and the colloidal 
glass transition. In the limit of vanishing thermal fluctuations the theory
becomes equivalent to our earlier theory for the glassy dynamics of the 
AOUP model (at the level of our description of the time evolution of the 
self-propulsion these models are equivalent). 
Finally, in the limit of rapidly varying self-propulsion the theory
becomes equivalent to the standard mode-coupling theory for the thermal system
at a temperature equal to the effective temperature. On the other hand, our
effective evolution operator (\ref{Omegaeff}) becomes ill-defined in the limit 
of very slowly evolving self propulsion, $D_r\to 0$. We note that 
the last limit is rather difficult since none of the present theories for the 
dynamics of dense active systems seem to work well in this limit, with the
exception of the theory of Liluashvili \textit{et al.}

According to the present theory, the location of the active ergodicity-breaking 
transition is determined by equal-time steady-state correlation functions only.
This is in contrast with the approach of Liluashvili \textit{et al.}, in which
the equation determining the ergodicity-breaking transition involves also 
a time integral over the decaying elements of the matrix correlator, which is the
basic fundamental quantity of that theory.

The main fundamental assumption behind the present theory is the vanishing 
of stead-state currents after integration over the self-propulsions. Preliminary
analytical results suggest that the presence of the currents may lead to wiping out
the ergodicity-breaking transition. A combined analytical and simulational 
study in this direction is planned for the near future. 

Finally, we note that, somewhat unexpectedly, the theory of Liluashvili \textit{et al.}
predicts that in addition to density correlations also some cross-correlations
between the density and the self-propulsion slow down and get arrested at the
ergodicity-breaking transition. While our theory cannot address such cross-correlation
functions directly, we feel that this is another area where a combined analytical
and simulational study would be very interesting. 

\section*{Acknowledgments}
I thank Elijah Flenner for comments on the manuscript.
I gratefully acknowledge the support of NSF Grant No.~CHE 1800282.



\begin{thebibliography}{99}
\bibitem{comment2d} The present theory could be generalized
to three spataial dimensions. In this case, the direction of the self-propulsion
velocity is described by two angular variables; the equation of motion for these 
variables is a bit more complicated than Eq. (\ref{emotion2}).
\bibitem{commentSP} The motion of swimming bacteria or self-propelled Janus
particles is believed to be force free but we follow previous studies \cite{tenHagen} 
and treat the self-propulsion as if it were an effective internal driving force.
\end{thebibliography}
\end{document}